\documentclass[12pt]{article}
\usepackage{graphicx}
\usepackage{amsmath}
\usepackage{lastpage}
\usepackage{moresize}
\usepackage{fancyhdr}
\begin{document}

{\bf A QUANTILE SHIFT APPROACH TO MAIN EFFECTS AND INTERACTIONS IN A 2-BY-2 DESIGN} 

\noindent
Rand R. Wilcox$^{1*}$ \& Guillaume A. Rousselet$^{2*}$\\ 
1. Dept. of Psychology, University of Southern California, Los Angeles, CA 90089-1061, USA
orcid.org/0000-0002-2524-2976\\
2. School of Psychology and Neuroscience, College of Medical, Veterinary and Life\\
Sciences, University of Glasgow, 62 Hillhead Street, G12 8QB, Glasgow, UK\\
orcid.org/0000-0003-0006-8729\\
$^*$ Corresponding authors: rwilcox@usc.edu, Guillaume.Rousselet@glasgow.ac.uk\\

\begin{abstract}
When comparing two independent groups, shift functions are basically techniques that  compare
multiple quantiles rather than a single measure of location, the goal being to get a more detailed 
understanding of how the distributions differ. Various versions have been proposed and
studied. This paper deals with extensions of  these methods to main effects and
interactions in a  between-by-between, 2-by-2 design. 
Two approaches are studied, one that compares the deciles of the distributions, and one that has a certain connection to the Wilcoxon--Mann--Whitney method. There are many 
quantile estimators, but for reasons summarized in the paper, the focus is on using the Harrell--Davis quantile estimator used in conjunction with a percentile bootstrap method. Included are results 
comparing two methods aimed at controlling the probability of one or more Type I errors.
\end{abstract}


\section{Introduction}

When comparing two distributions, certainly the most common approach is to focus on a single
measure of location, typically the mean or median. An alternative approach is to compare multiple quantiles with the goal of getting a more detailed understanding of where groups differ and by how much (Rousselet et al., 2017). Several methods have been derived for dealing with this issue  (e.g., Doksum \& Sievers, 1976; Goldman \& Kaplan, 2018; Lombard, 2005; Wilcox, 1995; Wilcox et al., 2014). 
Extant results suggest how to generalize these methods to 
 a between-by-between, 2-by-2 design. One goal here is to report results on two methods
 for controlling the family wise error rate (FWER), meaning the probability of one or more Type I errors.
 
 Here,  two distinct approaches are considered. The first defines interactions and main effects with means replaced by a collection of quantiles with an emphasis on the deciles. 
 For example, if $\theta_{jk}$ denotes the .2 quantile corresponding to level $j$ of the first factor and level $k$ of second factor, one goal is to test 
\begin{equation}
H_0: \theta_{11}-\theta_{12}=\theta_{21}-\theta_{22}, \label{eq:inter}
\end{equation}
which mimics the usual notion of an interaction in an obvious way.
Here, however, the goal is to use multiple quantiles and to assess how well the
FWER is controlled.
Of course, a related issue is computing a reasonably accurate confidence interval for
$\theta_{11}-\theta_{12}- \theta_{21}+ \theta_{22}$. 
As is evident, main effects can be addressed as well. For example, one can test
 \begin{equation}
H_0: \theta_{11}+\theta_{12}=\theta_{21}+\theta_{22}, \label{eq:main}
\end{equation}
for a collection of quantiles.

The second approach, when dealing with an interaction, has a certain connection
to  a rank-based method proposed by Patel and Hoel (1973) that in turn has a connection with 
 the classic Wilcoxon--Mann--Whitney test. To describe the Patel--Hoel approach, let $X_{jk}$ denote four independent random variables where $X_{11}$ and $X_{12}$ correspond to the first level of the first factor  in a 2-by-2 design, while $X_{21}$ and $X_{22}$ correspond to the second level of the first factor.
 Let $p_1 = P(X_{11} < X_{12})$ and $p_2 = P(X_{21} < X_{22})$. The hypothesis of  no interaction is 
 \begin{equation}
H_0: p_1 = p_2.
\end{equation}
And  there is the issue 
of computing a $1-\alpha$ confidence interval for $p_1-p_2$.
Wilcox (2022, section 7.9.2) describes a method for making inferences about this measure of effect size that performs well in simulations. For some related rank-based methods, see Gao and Alvo  (2005), as well as De Neve and Thas (2017).

Let $X_{i jk}$ denote a random sample from the $j$th level of the first factor and the
$k$th level of the second factor
($i=1, \ldots, n_{jk}$;  
$j=1$, 2; $k=1$, 2).
For convenience momentarily focus on $p_1$ and let $D_{i h} = X_{i11}-X_{h 12}$ ($i=1, \ldots, n_{11}$; $h=1, \ldots, n_{12}$).
An estimate of  $p_1$ is simply 
\begin{equation}
\hat{p}_1= \frac{1}{n_1 n_2} \sum \sum I(D_{i h}),
\end{equation}
where the indicator function $I(D_{i h})=1$ if $D_{i h}<0$; otherwise $I(D_{i h})=0$.
The estimator $\hat{p}_1$ is the estimator used by the classic Wilcoxon--Mann--Whitney
test.

Let $\theta_1$ denote the median of $D$ and notice that 
\begin{equation}
H_0: p_1= .5
\end{equation}
is the same as
\begin{equation}
H_0: \theta_1=0. 
\end{equation}

The parameter $\theta_1$ is defined based on level 1 of the first factor.
Let $\theta_2$ denote the analog of $\theta_1$ when dealing with level 2 of the first factor.
Then an analog of the Patel--Hoel interaction is
$\theta_1-\theta_2$. 
Here, however, the goal is to consider the broader issue of comparing the 
deciles of these two distributions. More formally, let 
$q_1$  and $q_2$ denote the $q$th quantile 
of the distribution $D$  for  level 1 of the first factor and level 2 of first
factor, respectively. The goal is to test
\begin{equation}
H_0: q_1=q_2 \label{eq:apd}
\end{equation}
and to compute a $1-\alpha$  confidence interval 
for $q_1-q_2$ for  $q=.1, .2, \ldots, .9$.
A second goal is to control the FWER in a reasonably accurate manner.

Notice that when testing (\ref{eq:inter}), there is no distinction between
$\theta_{11}-\theta_{12}=\theta_{21}-\theta_{22}$ and
$\theta_{11}-\theta_{22}=\theta_{12}-\theta_{22}$.
That is, interchanging the rows and columns does not alter the estimated effect size.
However, when dealing with (\ref{eq:apd}), interchanging the rows and columns can yield different results.

To illustrate this point, consider, for example, a situation where for the first level of the first factor, both levels of the second factor have standard normal normal distributions,
while for the second level of the first factor, the two levels of the second factor
have lognormal distributions. Data were generated as just described 
based on sample sizes of 50 per group and the 50th quantile was estimated. The estimate of $q_1-q_2$ was .027.
But interchanging the rows and columns, now the estimate was $-.088$. The code to reproduce this example can be found in the R notebook \verb|apd_ex.Rmd|, which is available as part of the companion reproducibility package for this article (Wilcox \& Rousselet, 2023) and on GitHub (Rousselet, 2023).



\section{The Proposed Methods}

First consider testing (\ref{eq:inter}). The first issue is choosing a reasonable quantile 
estimator from among the  many estimators that might be used. The focus here is on the
 estimator derived by Harrell and Davis (1982)
 for two fundamental reasons. The first has to do with tied values.
When comparing two independent groups using the usual sample median,
tied values can be accommodated using  a slight generalization of a standard
percentile bootstrap method. However, when comparing other quantiles using an
estimator based on only one or two order statistics, such as those 
summarized by   Hyndman and Fan (1996), this approach no longer performs
in an adequate manner (e.g., Wilcox, 2022).
Simulation results reported in section 3 indicate that  this remains the case for the
situation at hand.

In contrast to the estimators considered by Hyndman and Fan (1996), 
 the  Harrell--Davis estimator uses a weighted average of all the order statistics.
Moreover, when using the  Harrell--Davis estimator  
in conjunction with a percentile bootstrap method, all indications are that
this approach  does perform well when comparing two independent groups and 
there are tied values. An issue here is whether this continues to be the case when
dealing with a two-way design. 
A second reason for using  the Harrell--Davis estimator is that is has 
 better efficiency under normality compared to using a weighted average of only two order statistics.
 
  For completeness, it is noted that comparing quantiles can be accomplished using 
 the quantile regression estimator derived by Koenker and Bassett (1978).
 However, this is tantamount to using one or two order statistics. When comparing the medians
 of two independent groups, for example, and the sample sizes are odd, in effect the usual
 sample median is being used. Evidently, there is no generalization of  the Koenker--Bassett method
 that captures the spirit of the Harrell--Davis estimator.

 There are several quantile estimators, in addition to the Harrell--Davis
 estimator, that use all of the order 
 statistics (e.g., Liu et al., 2022; Navruz \&  \"{O}zdemir, 2020).
 A possible criticism is that these estimators, including the Harrell--Davis estimator,
  have a breakdown point of only 
 $1/n$. That is, the minimum number of order statistics that must be
 altered to make the estimate arbitrarily large is one.
 In practice this issue might not be a serious concern because the
 extreme order statistics get a relatively small weight. In a situation where
 the breakdown point is an issue, one possibility is to use the
 trimmed Harrell--Davis estimator derived by Akinshin (2022). 
 
\begin{figure*}[!ht]
\centering
\includegraphics[width=0.9\textwidth]{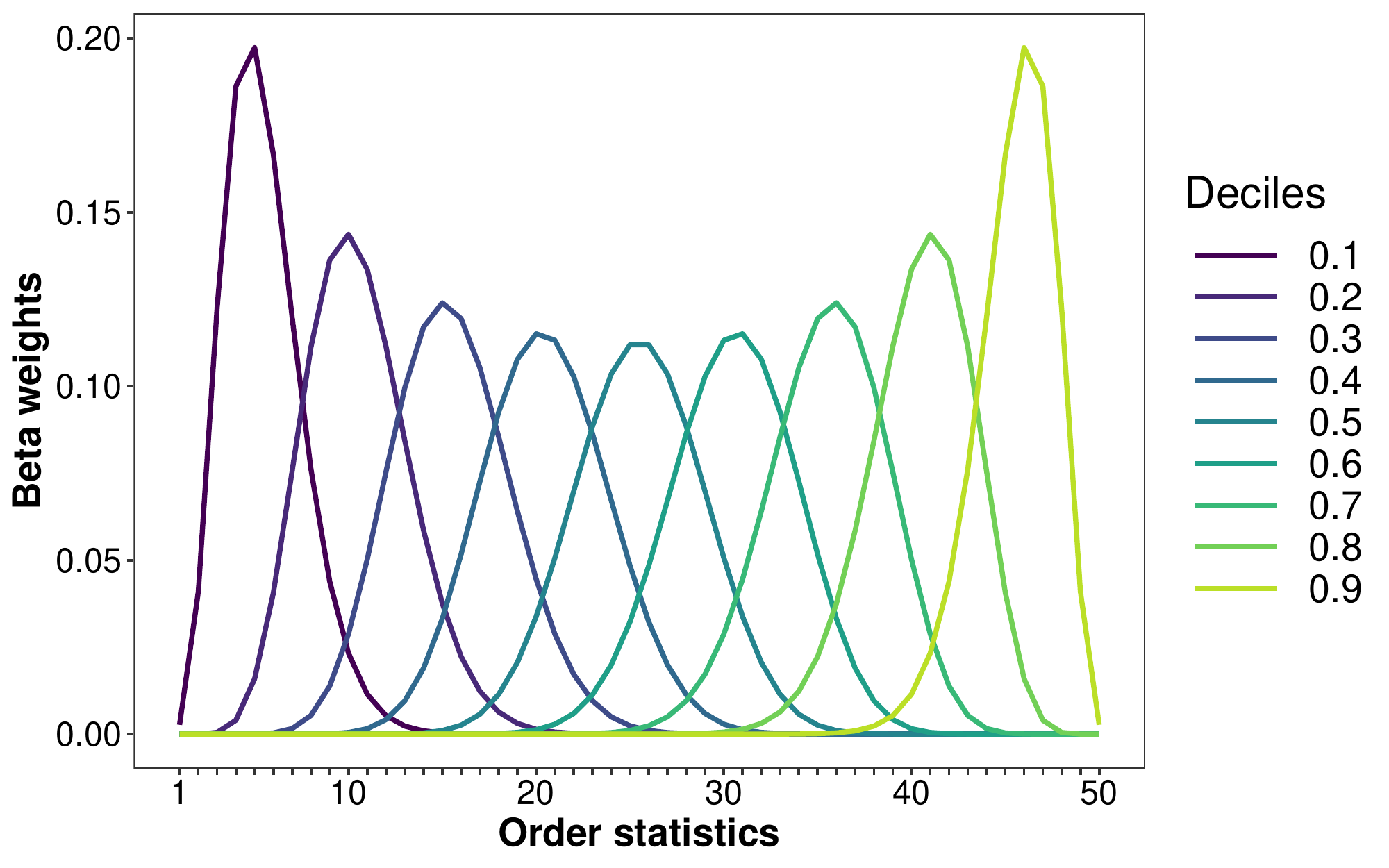}
\caption{
\begin{small}
\textbf{Beta weights used to calculate the Harrell-Davis estimates of the deciles with a sample size of 50.}
\end{small}}
\label{fig:betaweights}
\end{figure*}

Let $U$ be a random variable having a beta distribution with parameters
$a=(n+1)q$ and $b=(n+1)(1-q)$.  Let
\[W_i = P\left(\frac{i-1}{n} \le U \le \frac{i}{n}\right).\]
The Harrell--Davis estimate of the $q$th quantile is
\begin{equation}
\hat{\theta}_{q} = \sum_{i=1}^n W_i X_{(i)}, \label{hdest}
\end{equation}
where $X_{(1)} \le \cdots \le X_{(n)}$ are the values written in ascending order. The beta weights used to calculate the deciles for a sample size of $n=50$ are illustrated in Figure \ref{fig:betaweights}.

As an alternative to the Harrell--Davis estimator, we considered the default quantile estimator in R, called using the command \verb|quantile(x, probs=seq(0.1,0.9,0.1), type=7)|,  to compute the deciles. This estimator, in addition to being widely used, relies on two order statistics (Hyndman \& Fan, 1996), and could thus be more robust to outliers than the Harrell--Davis estimator in some situations.

To test (\ref{eq:inter}), we considered a percentile bootstrap method combined with the Harrell--Davis estimator and the quantile(type=7) estimator. The percentile bootstrap method begins by generating  bootstrap samples by sampling with replacement $n_{jk}$ values from the data associated with the $j$th level of the first factor and the $k$th level of second factor yielding  $X^*_{ijk}$ ($i=1, \ldots, n_{jk}$).
Based on these bootstrap samples, compute the $q$th quantile using the
Harrell--Davis estimator, or the quantile(type=7) estimator, yielding $\theta^*_{jk}$ followed by
\begin{equation}
\Psi^*=\theta^*_{11}-\theta^*_{12}- \theta^*_{21}+ \theta^*_{22}.
\end{equation}
Repeat this process $B$ times yielding $\Psi^*_1, \ldots, \Psi^*_B$.
Let $A$ denote the number of $\Psi^*$ values that are less than zero and let
$D$ denote the number of $\Psi^*$ values that are equal to zero.
Let 
\begin{equation}
P =  \frac{A}{B} +.5 \frac{D}{B}.
\end{equation}
A p-value, when testing (\ref{eq:inter}), is $2 \min\{P, 1-P\}$.
The term $D/B$ is important when dealing with
tied values (e.g., Wilcox, 2022).
To compute a $1-\alpha$ confidence interval, put 
$\Psi^*_1, \ldots, \Psi^*_B$ in ascending order yielding 
$\Psi^*_{(1)} \le  \cdots \le  \Psi^*_{(B)}$.
Let $\ell= \alpha B/2$, rounded to the nearest integer. Let $u=n- \ell$. 
Then a  $1-\alpha$ confidence interval for $\Psi$ is
\begin{equation}
(\Psi^*_{(\ell +1)}, \Psi^*_{(u)}).
\end{equation}
Here, $B=2000$ is used. 

Note that the same bootstrap samples were used for each of the
quantiles being compared. An alternative approach is to use separate
bootstrap samples for each test to be performed (Wilcox, 1995). Both approaches were
considered and there is no indication that separate bootstrap samples offer
a practical advantage in terms of controlling the Type I error 
probability. Results of the simulations comparing the two bootstrap approaches are available in the R notebooks \verb|sim_fp_b1b9.Rmd| and \verb|sim_fp_apd_b1b9.Rmd| (Wilcox \& Rousselet, 2023). Using the same bootstrap samples for all of the tests performed considerably reduces execution time, which is why it is assumed henceforth. 

When performing $C$ tests, there is the issue of controlling the FWER, meaning the probability of one or more Type I errors. Two approaches are considered here. The first is
Hochberg's (1988)  improvement on the Bonferroni method. Let  $p_1, \dots, p_C$ be the
p-values associated with the $C$ tests. Put these p-values in descending order, and label the results
$p_{[1]} \ge p_{[2]} \ge \cdots \ge p_{[C]}$.
Set $k=0$ 
and proceed as follows:

\begin{enumerate}

\item
Increment $k$ by 1.
If
\[p_{[k]} \le  \frac{\alpha}{k},\]
 stop and reject all hypotheses having
a p-value less than or equal to $p_{[k]}$.

\item If $p_{[k]} > \alpha/k$, repeat step 1.

\item
Repeat steps 1 and 2 until a significant result is obtained or all $C$ hypotheses have  been tested.
\end{enumerate}

The second method for controlling the FWER was Benjamini and Hochberg (1995), which is aimed at controlling the false discovery rate (FDR). That is, the goal is to control the expected proportion
of Type I errors among the null hypotheses that are correct. It is known that there are situations where 
 the Benjamini--Hochberg method ensures that the false discovery rate is less than
 or equal to $\alpha$, but it does not necessarily control the FWER (Hommel, 1988).
 However, the simulations described in section 3, below, indicated that
 when using Hochberg's method, the FWER is always below the
 nominal level. Consequently, there was interest in how well the
 Benjamini--Hochberg  method performs. It is readily verified that
 the Benjamini--Hochberg method always has as much or more power than Hochberg's method.
Consequently, provided the Benjamini--Hochberg method
controls the FWER for the situation at hand, it has a practical advantage over Hochberg's method.
 
 The Benjamini--Hochberg method is applied simply by replacing 
 $p_{[k]} \le \alpha/k$ in step 1 of Hochberg's method with
\begin{equation}
p_{[k]} \le  \frac{(C-k+1)\alpha}{C}
\end{equation}

As for testing (\ref{eq:apd}), again both the Harrell--Davis and the quantile(type=7) estimators
 were considered, in conjunction with a percentile bootstrap method. 
 Bootstrap samples are generated as before yielding 
 estimates of $q_1$ and $q_2$, which are labelled
$q^*_1$ and $q^*_2$. This process is repeated $B$ times and the results are
used to compute a $1-\alpha$ confidence interval for  $q_1 -q_2$, as well
as a  p-value, in essentially the same manner as done when testing (\ref{eq:inter}).

\section{Simulation Results}


Simulations were used to check the FWER when testing 
(\ref{eq:inter}) and (\ref{eq:apd}),
as well as how the power of these methods compare to the classic ANOVA
F test as well as a method for comparing 20\% trimmed mean, which
is described in Wilcox, 2022, section 7.4.3).
Data were generated from four continuous distributions as well as three discrete distributions. The discrete distributions were a Poisson distribution having mean 9, and two beta-binomial distributions, one with parameter $r=1$, the other with $r=9$, and the other parameters set to $s=9$, and 10 bins. The three discrete  distributions were included as a partial check on the impact of tied values.
The four continuous distributions
were  standard normal,
 mixed normal, lognormal and mixed lognormal.
 The distribution of the mixed normal is
\begin{equation}
 H(x)= .9 \Phi(x) + .1 \Phi(x), \label{cn:dist}
\end{equation}
where $\Phi(x)$ is the standard normal distribution. 
The mixed normal is a symmetric distribution with heavy tails, roughly meaning that outliers are likely to occur.
The mixed lognormal distribution is given by (\ref{cn:dist}), but with
$\Phi$ replaced by the lognormal distribution. 

Based  on over 1,500 estimates of skewness and kurtosis
reported in various journal articles, Cain et al. (2017) report that 
99\% of the estimates were less than the skewness and 
kurtosis of a lognormal distribution. This suggests that if a method 
performs reasonably when dealing with a lognormal distribution, it is 
highly likely to perform reasonably well in practice. 
However, a possible concern is that when dealing with heavy-tailed 
distributions, the standard error of the usual kurtosis estimator can be
quite high even when the sample size is fairly large. 
Moreover, the usual estimate of kurtosis can grossly
under estimate the true value.

Consider, for example, a
lognormal distribution, which has kurtosis 113.9.
Based on a sample of 100,000, the kurtosis of the lognormal distribution was
estimated and this
process was repeated 1000 times. 
It was found that 79\% of the estimates were less than the true value.
The median estimate was 82.
This process was repeated using the mixed lognormal distribution which
is skewed and very heavy-tailed. The kurtosis of this distribution
was estimated based on a sample size of one million. This process was
repeated 1000 times yielding estimates ranging between 242 and 16400.
The median estimate was 429. Consequently, the
mixed lognormal distribution was used here as an
additional check on how well the methods perform. The code for these simulations is available in the notebook \verb|kurtosis_estimation.Rmd| (Wilcox \& Rousselet, 2023).



Simulations were performed for sample sizes 20, 30, 40, ..., 100 per group, using 10,000 iterations.
This was done using both types of quantile estimators and for the two main effects and the interaction. A p-value was computed for each of the nine quantiles to be compared. 
In terms of controlling the FWER, both Hochberg's method and the
Benjamini--Hochberg method were considered.
And as previously indicated, this was done for seven distributions.
So in terms of Type I errors 
there is a total of 3402 results (7 distributions * 2 quantile estimators * 3 contrasts * 9 deciles * 9 sample sizes). 

Finally, simulations were performed dealing with power. Various situations were considered, including shifting the four continuous distributions used to estimate Type I errors, varying the skewness of \emph{g-and-h} distributions (Hoaglin, 1985), and the skewness of Poisson and beta-binomial distributions. In all situations, the distributions were adjusted to provide high power at the maximal sample size, to better differentiate the various methods. Complete details, including the code that was used, are
reported in files available on figshare (Wilcox \& Rousselet, 2023) and on GitHub (Rousselet, 2023). The main R packages used to perform the simulations and illustrate the results were 
\verb|Rcpp| (Eddelbuettel \& Francois, 2011), \verb|ggplot2| (Wickham, 2016) and \verb|cowplot| (Wilke, 2017).

First we consider the results for testing (\ref{eq:inter}), then for testing (\ref{eq:apd}).

\subsection{ Results when testing 
(\ref{eq:inter})}

The results regarding the FWER were highly consistent over the conditions considered, making it a simple matter to briefly summarize the relative merits of the methods being considered. All indications are that the actual FWER is less than the nominal .05 level using  the Hochberg method as well as the Benjamini--Hochberg method.
Consequently, the Benjamini--Hochberg method is recommended because it always
has as much or more power than Hochberg's method.

\subsubsection{Type I errors}

\begin{figure*}[!ht]
\centering
\includegraphics[width=1\textwidth]{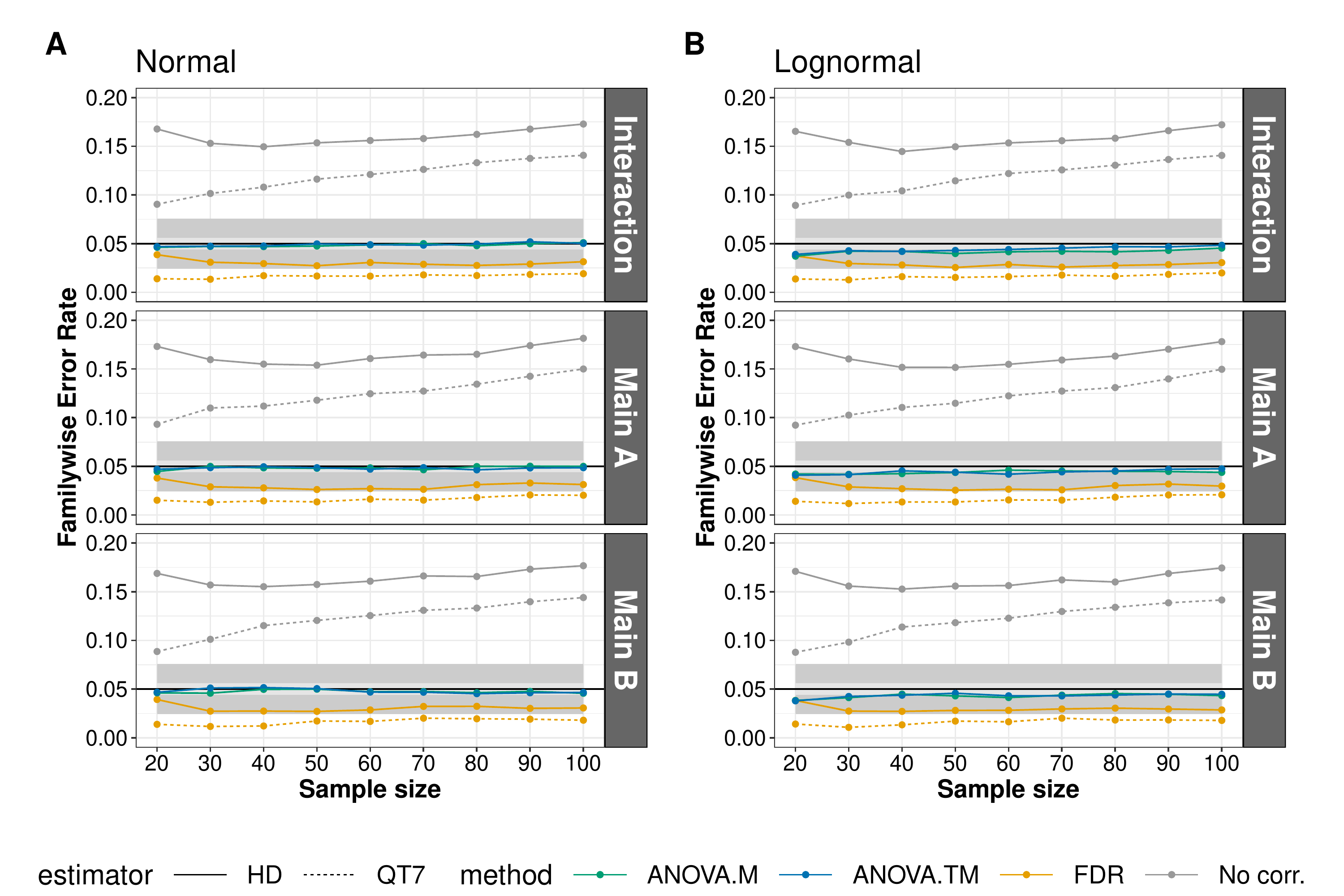}
\caption{
\begin{small}
\textbf{False positive results for normal and lognormal populations.} FWER (across quantiles) are plotted as a function of sample size per group for the interaction and the two main effects A and B. ANOVA.M = ANOVA using means. ANOVA.TM = ANOVA using 20\% trimmed means. FDR (Benjamini--Hochberg method) and No corr. (no correction) refer to the percentile bootstrap method in conjunction with the Harrell--Davis estimator (HD) and the quantile(type=7) estimator (QT7). Horizontal ribbons: dark grey indicates Bradley's (1978) satisfactory range [0.025; 0.075]; light grey indicates Bradley's (1978) ideal range [0.045; 0.055].
\end{small}
}
\label{fig:fp_norm_lnorm}
\end{figure*}

\begin{figure*}[!ht]
\centering
\includegraphics[width=1\textwidth]{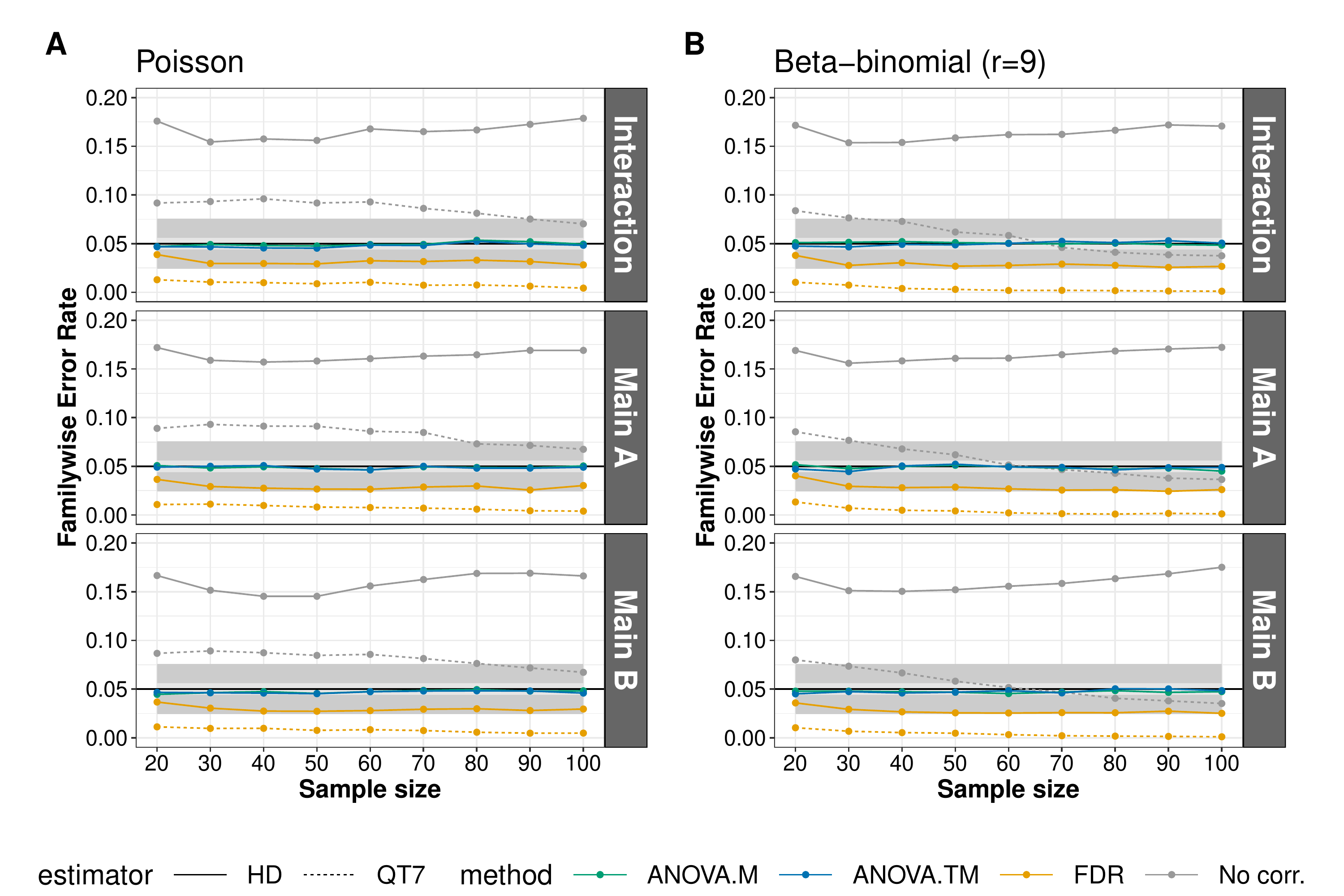}
\caption{
\begin{small}
\textbf{False positive results for Poisson and one of the two beta-binomial populations.} See details in the Figure \ref{fig:fp_norm_lnorm} caption.
\end{small}
}
\label{fig:fp_pois_bbr9}
\end{figure*}

To illustrate some of the type I error simulation results, Figure \ref{fig:fp_norm_lnorm} contains results for the normal and lognormal populations, separately for the interaction and the two main effects. Results with Hochberg's method are omitted but detailed figures are available in the notebook \verb|sim_fp.Rmd|. The FWER without correction for multiple comparison (grey lines) is lower than expected if the 9 tests were independent, which in theory would be $1-(1-\alpha)^9=0.37$. The results in \ref{fig:fp_norm_lnorm} and \ref{fig:fp_pois_bbr9} are about half of that expected value, supporting the use of the FDR correction over Hochberg's.

As a sanity check, results for the two types of ANOVAs are included, confirming false positives very close to the nominal level when sampling from a normal population. Although the percentile bootstrap led to conservative estimates for both quantile estimators, the Harrell--Davis estimator clearly outperforms quantile(type=7).

The gap between the two estimators increases when we consider samples from discrete populations, because the performance of quantile(type=7) deteriorates while that of Harrell--Davis remains stable (Figure \ref{fig:fp_pois_bbr9}).

Bradley (1978) suggested, as a general guide, that when testing at the .05 level, 
the actual level should be between .025 and .075.
This criterion was met in most situations for the Harrell--Davis estimator but not for the quantile(type=7) estimator.

\begin{figure*}[!ht]
\centering
\includegraphics[width=1\textwidth]{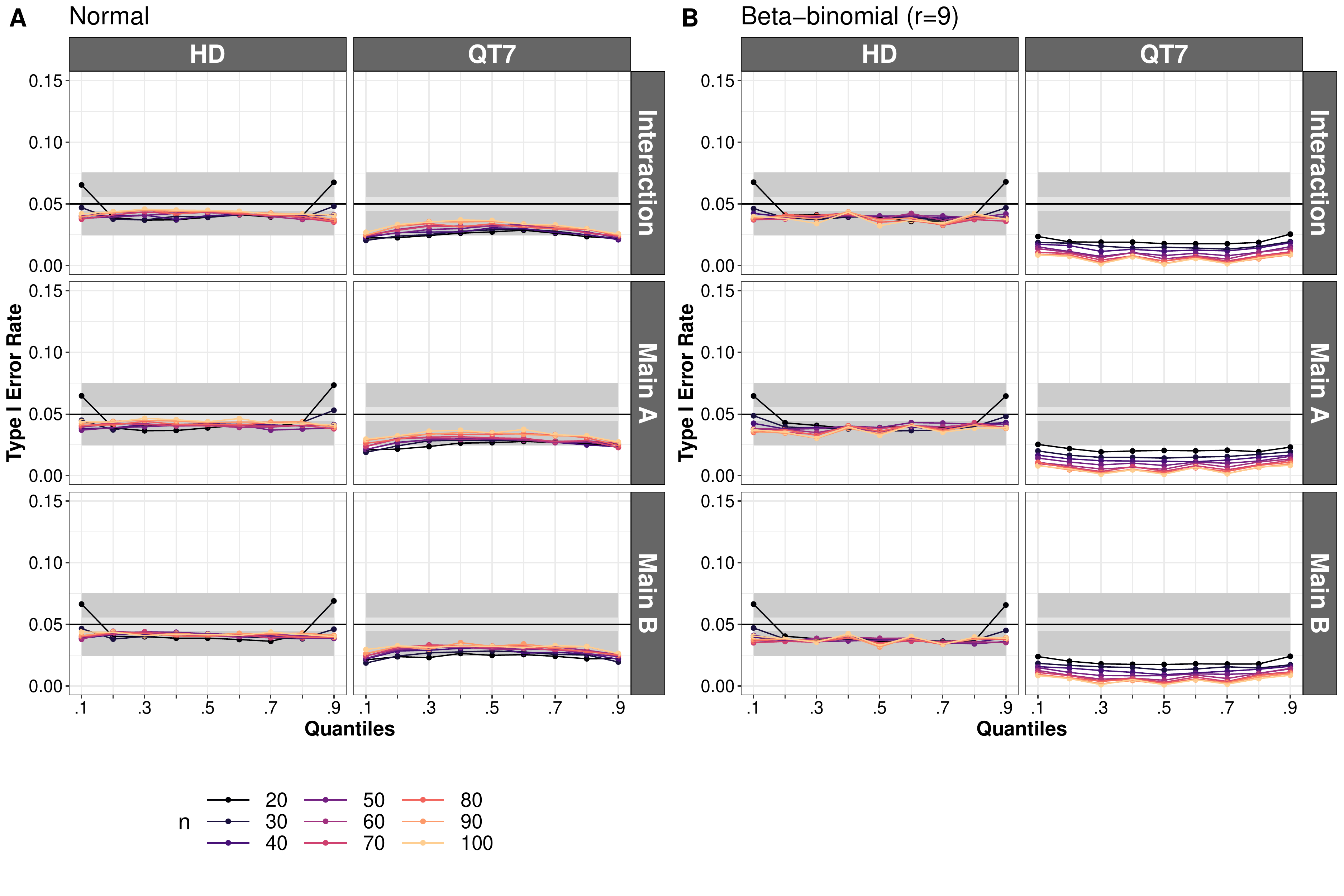}
\caption{
\begin{small}
\textbf{False positive results at individual deciles for normal and beta-binomial populations.}  Type I error rates are plotted at each decile, separately for the different sample sizes per group (n) and for the Harrell--Davis estimator (HD) and the quantile(type=7) estimator (QT7). Horizontal ribbons: dark grey indicates Bradley's (1978) satisfactory range [0.025; 0.075]; light grey indicates Bradley's (1978) ideal range [0.045; 0.055].
\end{small}
}
\label{fig:fp_ind_norm_bbr9}
\end{figure*}

The higher performance of the Harrell-Davis estimator compared to the quantile(type=7) estimator can be observed at individual deciles as well (Figure \ref{fig:fp_ind_norm_bbr9}). However, for the first and last deciles, the type I error rate is higher than the nominal level when using the Harrell--Davis estimator and n=20, a result that confirms earlier observations (Wilcox et al., 2014).  

\subsubsection{Power}
\begin{figure*}[!ht]
\centering
\includegraphics[width=1\textwidth]{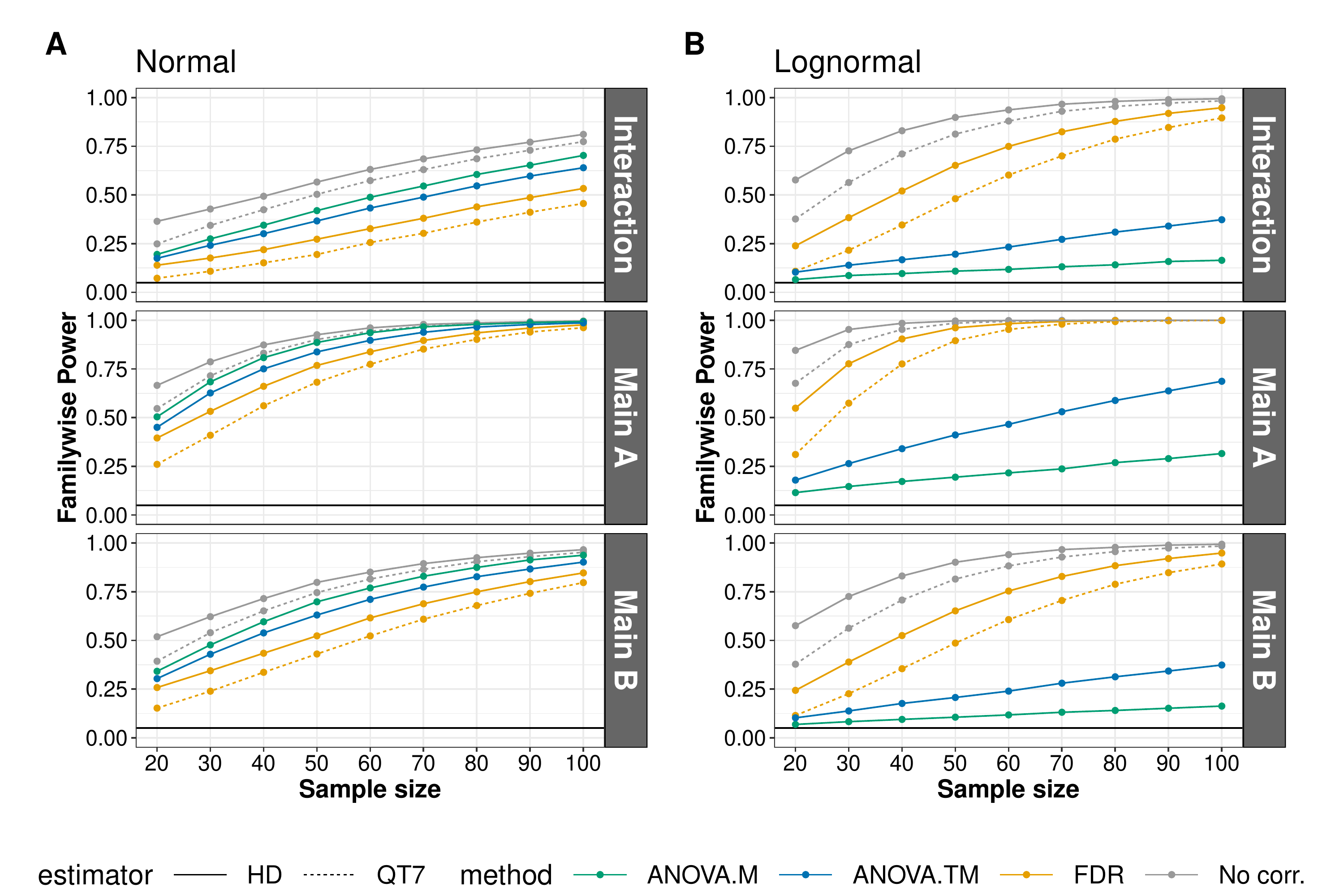}
\caption{
\begin{small}
\textbf{Power results for normal and lognormal populations.} ANOVA.M = ANOVA using means. ANOVA.TM = ANOVA using 20\% trimmed means. FDR (Benjamini--Hochberg method) and No corr. (no correction) refer to the percentile bootstrap method in conjunction with the Harrell--Davis estimator (HD) and the quantile(type=7) estimator (QT7). The black horizontal line marks 0.05. 
\end{small}
}
\label{fig:tp_norm_lnorm}
\end{figure*}

Not surprisingly, there are situations where inferences based on 
means or a 20\% trimmed have more power. But there are situations where
comparing deciles provides more power: no single method dominates. To illustrate, Figure \ref{fig:tp_norm_lnorm} presents results for normal and lognormal populations. In each case, data were generated for the four groups by sampling from the standard normal and lognormal distributions, before shifting the groups by different amounts. When sampling from normal populations, the ANOVA on means dominates other methods. To compare methods, we report familywise power for the decile methods (bootstrap p values with and without FDR correction), the probability of at least one rejection among the 9 tests. The ANOVA on 20\% trimmed means is a bit less powerful, followed by the bootstrap method using the Harrell--Davis estimator, and last the bootstrap method combined with the quantile(type=7) estimator. Switching to lognormal populations, the power of both ANOVA tests is dramatically lower than the bootstrap approach. This figure and the next one were generated using the notebook \verb|sim_tp.Rmd|.

\begin{figure*}[!ht]
\centering
\includegraphics[width=1\textwidth]{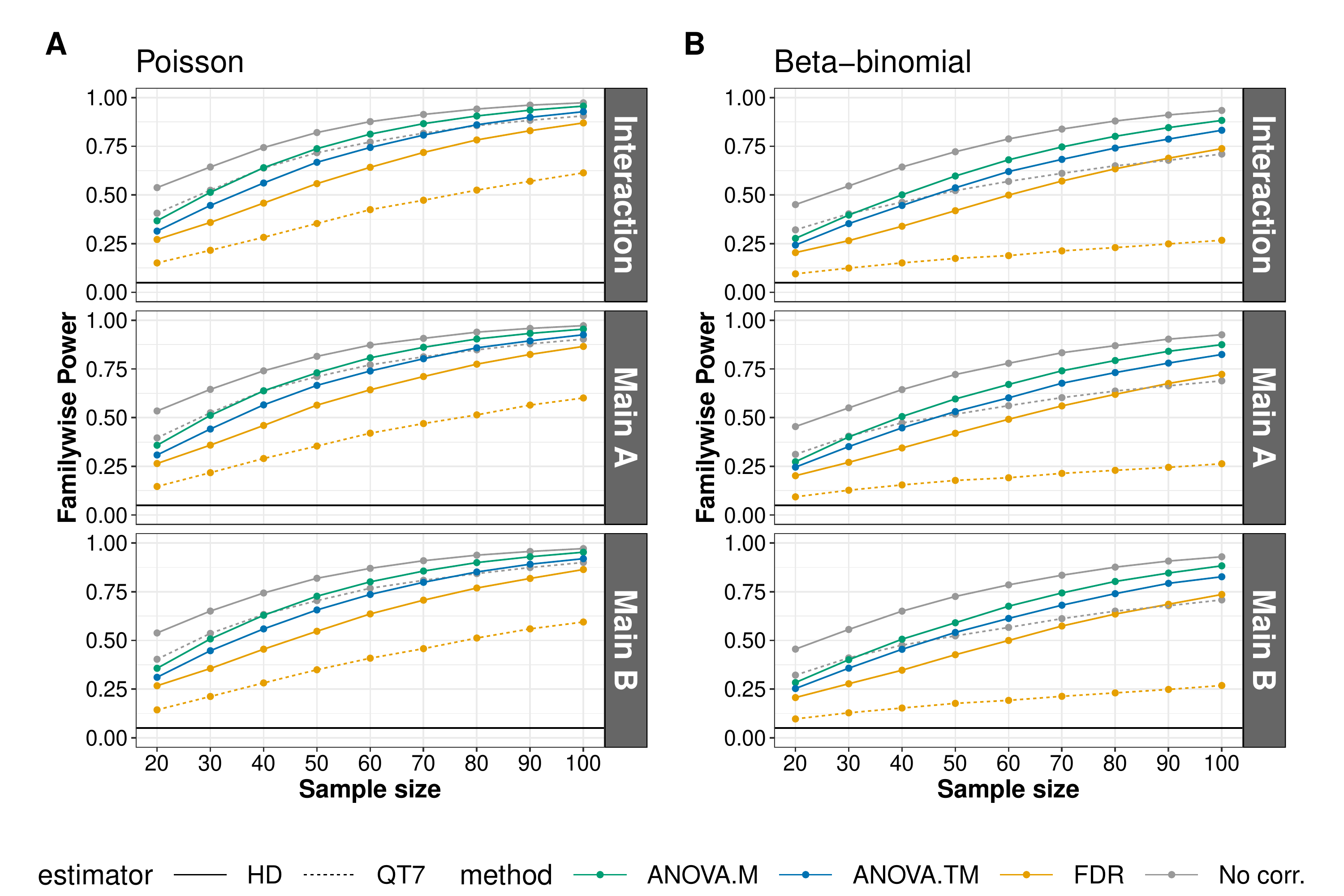} 
\caption{
\begin{small}
\textbf{Power results for Poisson and beta-binomial populations.} See details in Figure \ref{fig:tp_norm_lnorm} caption. 
\end{small}
}
\label{fig:tp_pois_bb}
\end{figure*}

Figure \ref{fig:tp_pois_bb} presents results from populations in which tied values were common. In both cases, the ANOVA on means dominates other methods. When using the bootstrap approach, the gap between the Harrell--Davis estimator and the quantile(type=7) estimator is larger than what was observed when sampling from continuous populations. 

\subsection{Compare deciles of distributions of all pairwise differences -- test \eqref{eq:apd}}

To assess the bootstrap method aimed at testing \eqref{eq:apd}, we used the same approach employed in the previous section. Now only the interaction is considered. The simulations and illustrations of the type I error rates can be found in the notebook \verb|sim_fp_apd.Rmd|. For power, see notebook \verb|sim_tp_apd.Rmd|. 

\subsubsection{Type I errors}

\begin{figure*}[!ht]
\centering
\includegraphics[width=0.95\textwidth]{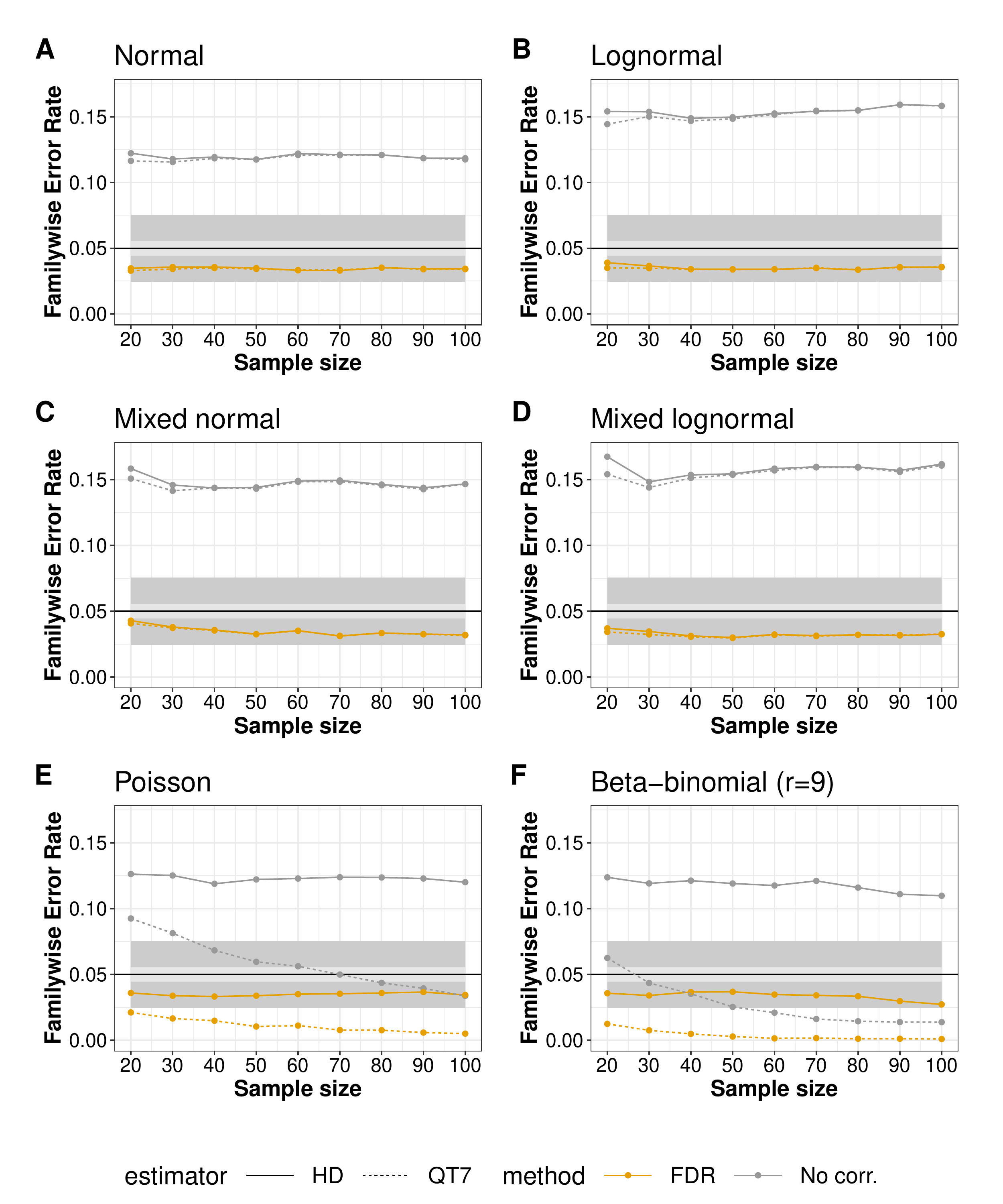} 
\caption{
\begin{small}
\textbf{FWER results for the comparison of the deciles of distributions of all pairwise differences.} 
\end{small}
}
\label{fig:fp_apd}
\end{figure*}

Again we observed FWERs much lower than expected if the deciles were independent (Figure \ref{fig:fp_apd}). For continuous distributions (panels A-D in Figure \ref{fig:fp_apd}), results were very similar for the Harrell--Davis and quantile(type=7) estimators. Both methods were conservative. When sampling from distributions in which tied values are likely, now the Harrell--Davis estimator outperforms the quantile(type=7) method (panels E and F in Figure \ref{fig:fp_apd}). 

\begin{figure*}[!ht]
\centering
\includegraphics[width=0.95\textwidth]{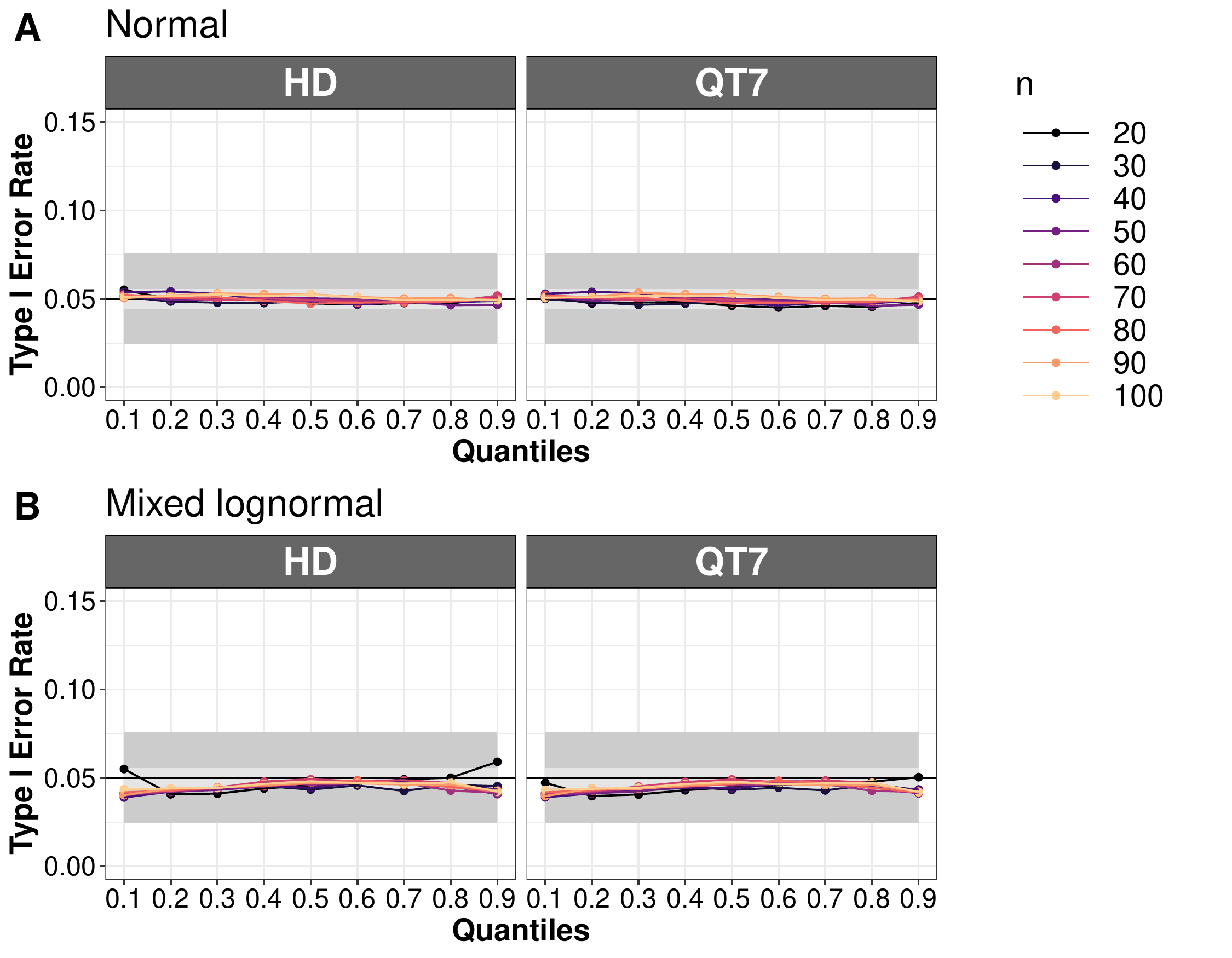} 
\caption{
\begin{small}
\textbf{Type I error rates for the comparison of the deciles of distributions of all pairwise differences: continuous distributions.}
\end{small}
}
\label{fig:fp_apd_ind_n_cln}
\end{figure*}

For continuous distributions, the similarity in performance between the two quantile methods is evident at the level of individual deciles as well (Figure \ref{fig:fp_apd_ind_n_cln}). Under normality, all deciles were associated with type I error rates close to the nominal level, irrespective of the sample size per group (panel A in Figure \ref{fig:fp_apd_ind_n_cln}). In the worst situation tested, when sampling from a mixed lognormal, results are a bit conservative, especially for the extreme deciles. In both situation, there is very little separating the two quantile methods.

\begin{figure*}[!ht]
\centering
\includegraphics[width=1\textwidth]{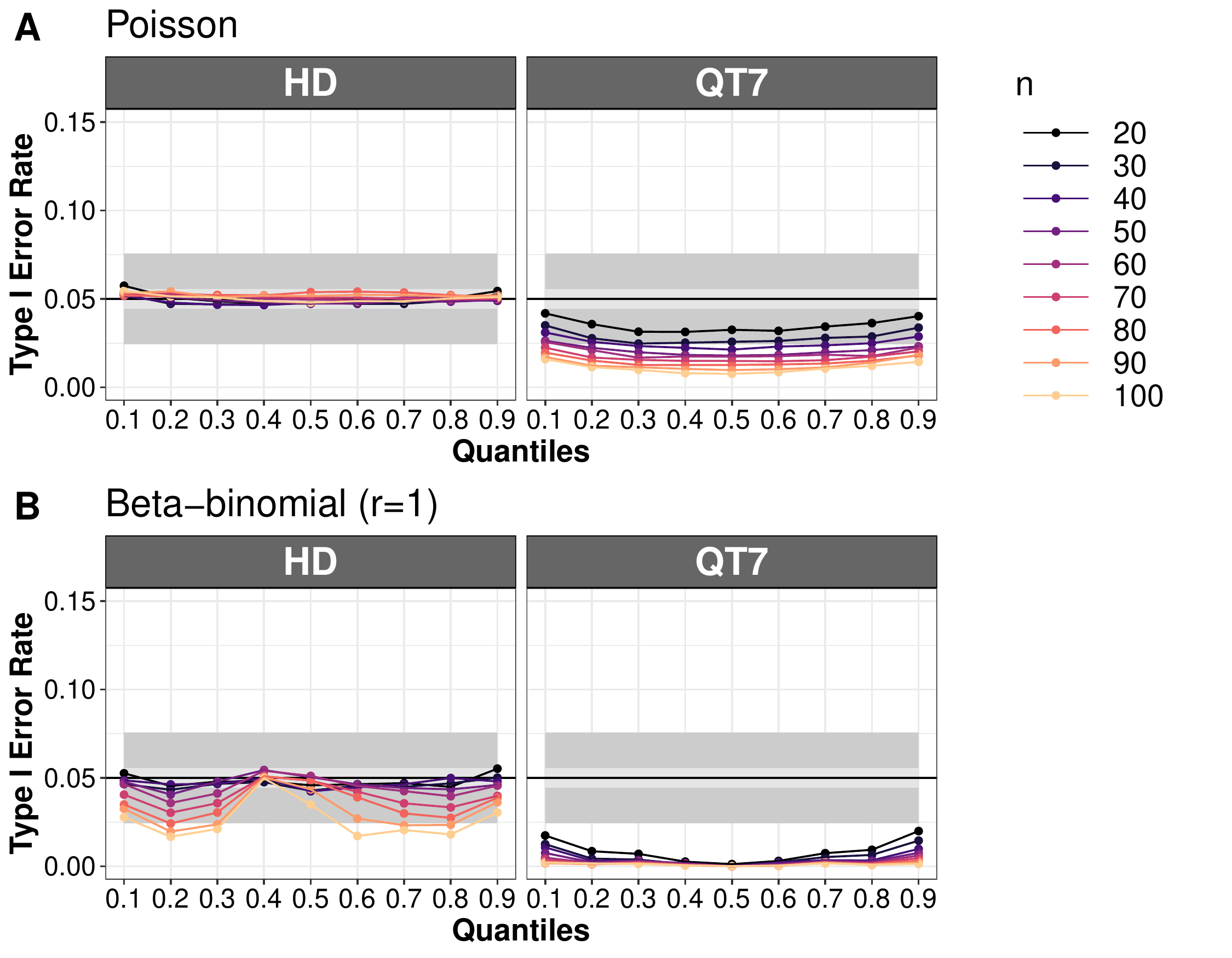} 
\caption{
\begin{small}
\textbf{Type I error rates for the comparison of the deciles of distributions of all pairwise differences: discrete distributions.}
\end{small}
}
\label{fig:fp_apd_ind_p_bb}
\end{figure*}

When sampling from a Poisson distribution (mean = 9 in all groups), the type I error rates for the Harrell--Davis estimator remain near 0.05, irrespective of sample size (panel A in Figure \ref{fig:fp_apd_ind_p_bb}). However, the quantile(type=7) is conservative and the situation deteriorates with increasing sample size. In the most extreme situation considered, when sampling from a beta-binomial distribution with $r=1, s=9, nbin=10$, the type I error rates were lower for both quantile estimators relative to the Poisson case (panel B in Figure \ref{fig:fp_apd_ind_p_bb}), or when sampling from a beta-binomial distribution with $r=9$ (not illustrated here, but see notebook \verb|sim_fp_apd.Rmd|). Although the situation got worse with increasing sample size for both estimators, Harrell--Davis outperformed quantile(type=7) in all situations.

\subsubsection{Power}

\begin{figure*}[!ht]
\centering
\includegraphics[width=0.95\textwidth]{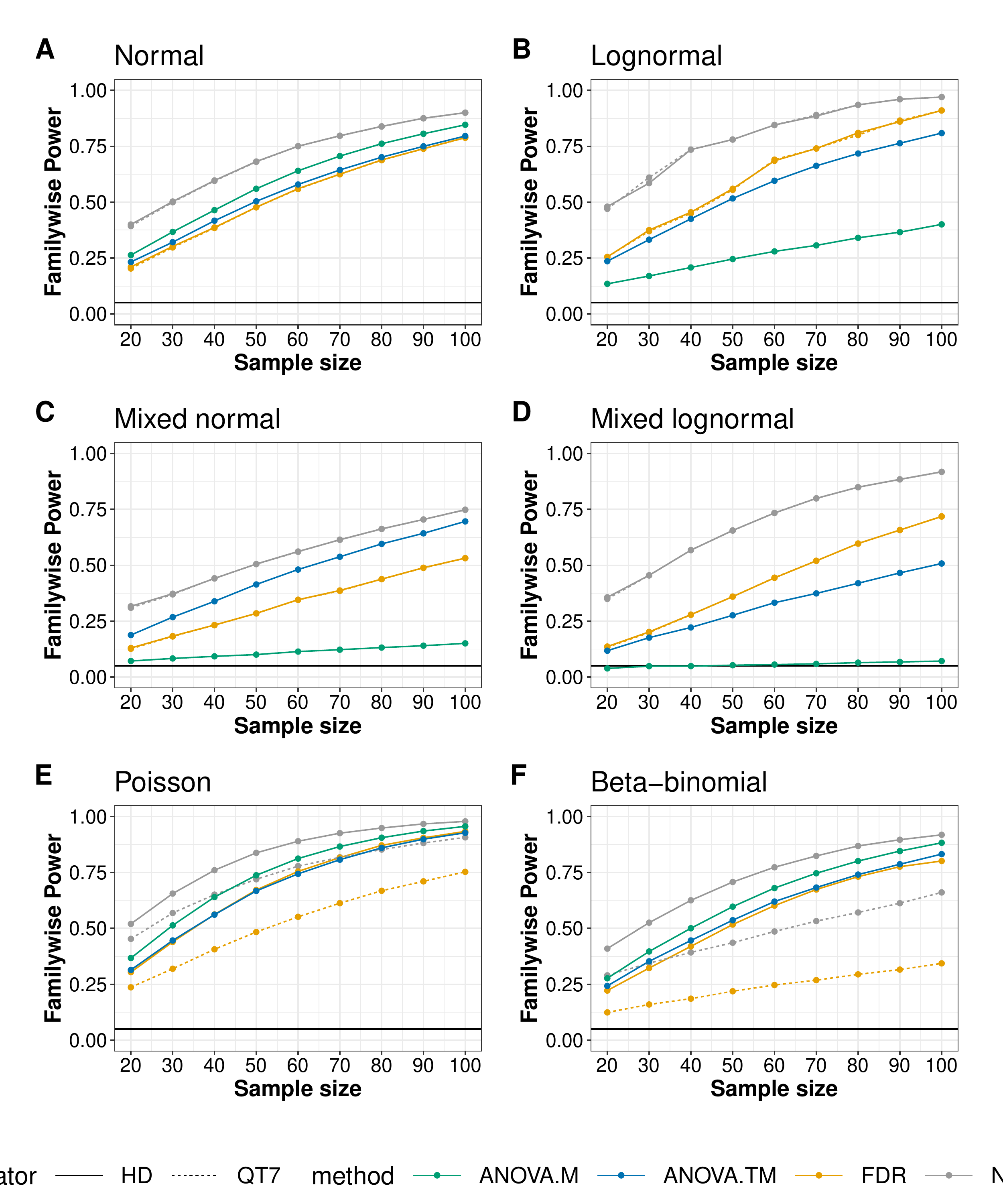} 
\caption{
\begin{small}
\textbf{Power results for the comparison of the deciles of distributions of all pairwise differences.} See details in Figure \ref{fig:tp_norm_lnorm} caption. 
\end{small}
}
\label{fig:tp_apd}
\end{figure*}

Under normality, the ANOVA on means performed best, followed by ANOVA on trimmed means and finally the bootstrap method (panel A in Figure \ref{fig:tp_apd}). When sampling from lognormal distributions, power was low for the ANOVAs relative to the bootstrap method, and much more so when making inferences about means (panel B in Figure \ref{fig:tp_apd}). For the mixed normal distributions, again the ANOVA on means performed poorly, but now the ANOVA on trimmed means dominates the bootstrap approach (panel C in Figure \ref{fig:tp_apd}). If sampling from mixed lognormal distributions, now the bootstrap method is the most powerful (panel D in Figure \ref{fig:tp_apd}). In all these situations, the Harrell--Davis and quantile(type=7) estimators gave very similar results. Finally, in the presence of tied values, the ANOVA on means dominated the other approaches, and the Harrell--Davis estimator led to higher power than the quantile(type=7) estimator (panels E and F in Figure \ref{fig:tp_apd}).

\section{An Illustration}

Both methods are illustrated using data dealing with perceived health (PH)
among older adults (Clark, et al., 2011). The first factor consists of two educational 
groups: those who did not complete high school and those who have some college or 
technical training. The other two groups are based on a measure of
depressive symptoms (CESD). One group corresponds to participants
with a CESD score greater than 15, which is often taken to indicate mild
depression or worse. The other level consists of participants with
CESD scores less than or equal to 15. The four groups are defined like this:

\begin{itemize}
  \item $A_1B_1$ = lower education, lower CESD, 
  \item $A_1B_2$ = lower education, higher CESD, 
  \item $A_2B_1$ = higher education, lower CESD, 
  \item $A_2B_2$ = higher education, higher CESD. 
\end{itemize}

\begin{figure*}[!ht]
\centering
\includegraphics[width=1\textwidth]{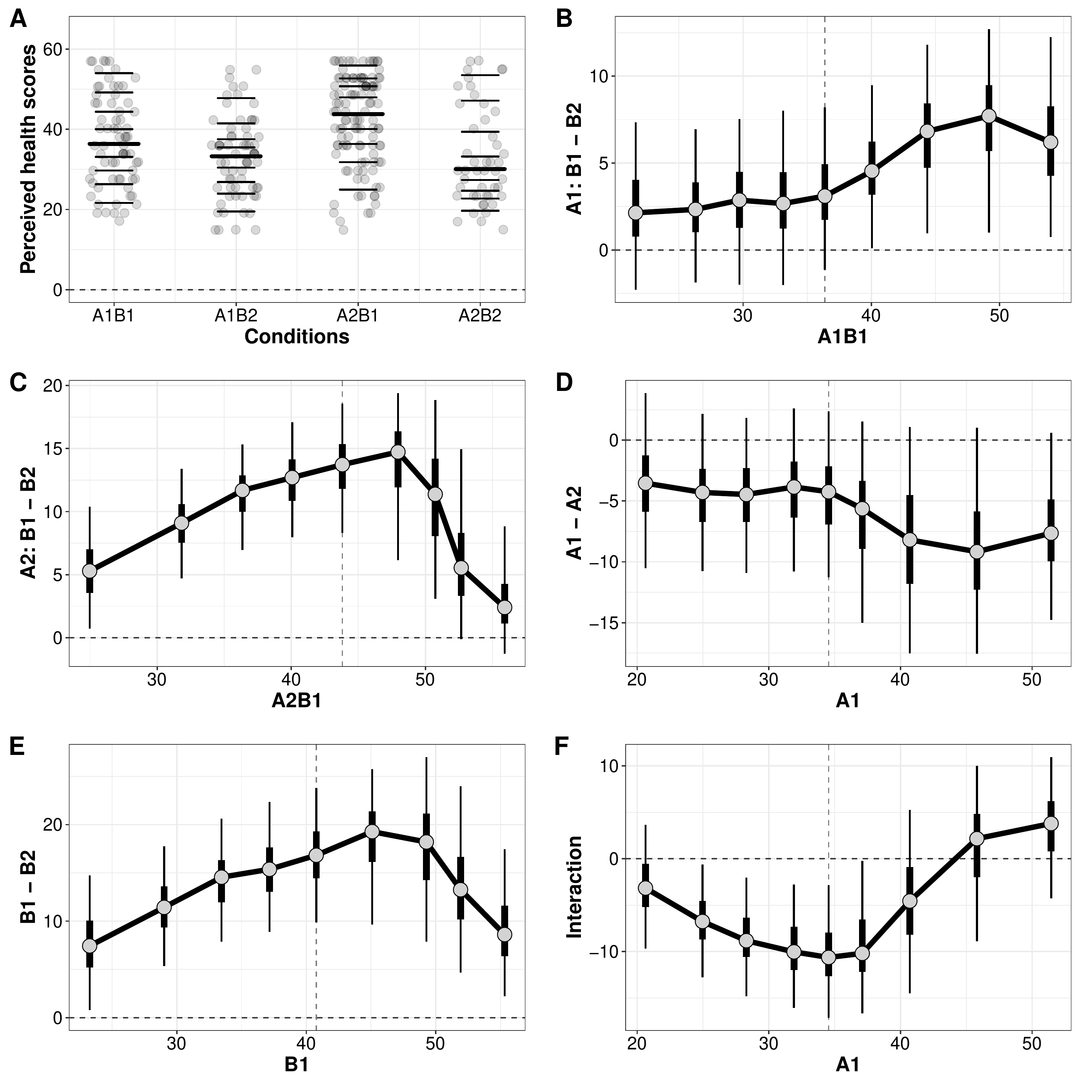} 
\caption{
\begin{small}
\textbf{Example: comparison of the deciles of perceived health.} (\textbf{A}) Perceived health scores in the 4 groups, with superimposed deciles indicated by horizontal lines. Medians appear as thicker lines. (\textbf{B}-\textbf{C}) Comparisons between B1 and B2 (lower/higher CESD) at each level of A. (\textbf{D}-\textbf{E}) Main effects. (\textbf{F}) Interaction. Grey disks: decile differences plotted as a function of the deciles in one group. Thick vertical lines mark 50\% confidence intervals. Thin vertical lines mark 95\% confidence intervals. Vertical dashed lines mark medians.
\end{small}
}
\label{fig:ex_dec}
\end{figure*}

Perceived health results are illustrated for the four groups in Figure \ref{fig:ex_dec}A. The figure was generated using the notebook \verb|examples.Rmd|. 

A 2x2 ANOVA on means returns these p values: main effect of A (education) = 0.001; main effect of B (depression) $<$ 0.0001; interaction = 0.09. Should we conclude that we have failed to obtain sufficient evidence about the presence of an interaction? This conclusion would be appropriate if the populations were symmetric and differed only in central tendency. However, the plot of marginals suggests differences in skewness and spread (Figure \ref{fig:ex_dec}A). In keeping with this observation, considering the deciles reveals a more complex picture, as illustrated in panels B-F of Figure \ref{fig:ex_dec}, with patterns of non-uniform group differences. Panels B and C illustrate the shift functions comparing B1 and B2 at each level of A: the decile differences between two groups are plotted as a function of the deciles in one group. Panels D and E illustrate the main effects. The values along the x axis (A1 and B1) correspond to the deciles of observations pooled across groups (for instance A1 = (A1B1, A1B2). Computing the average of the deciles leads to very similar graphs. Finally, panel F illustrates the interaction, which is highly non-linear, growing from the first decile to the median, and then decreasing and reversing sign. 


The output of the \verb|decinter| R function that tests (\ref{eq:inter}) is shown in
Table 1. 
As can be seen, the unadjusted p-values suggest that there is an interaction for the .2-.6 quantiles.
Shown are the adjusted p-values based on Hochberg's method
(using the R function \verb|p.adjust|) in order to underscore the
practical advantage of the Benjamini--Hochberg method.
As indicated, no significant difference is found at the .05 level using  Hochberg's method.
But using  the Benjamini--Hochberg correction when testing (\ref{eq:inter}), 
 the adjusted p-values, when dealing with the .3, .4 and .5 quantiles are all equal to .030.

\begin{table}
\center
\label{test1}
\caption{Result for perceived health when testing (\ref{eq:inter})}
\begin{verbatim}
     Quant Est.Lev 1 Est.Lev 2        Dif     ci.low      ci.up p-value p.adj
 [1,]   0.1  2.137558  5.295364  -3.157806  -9.898611  3.9567186   0.389 0.775
 [2,]   0.2  2.334238  9.091485  -6.757246 -13.234037 -0.3524981   0.039 0.195
 [3,]   0.3  2.865681 11.685416  -8.819735 -15.196697 -2.2297824   0.010 0.070
 [4,]   0.4  2.670866 12.699564 -10.028698 -16.657294 -2.9629910   0.008 0.070
 [5,]   0.5  3.098901 13.724232 -10.625331 -17.541216 -3.2211028   0.009 0.070
 [6,]   0.6  4.539717 14.729066 -10.189350 -16.841309 -0.8789187   0.036 0.195
 [7,]   0.7  6.823277 11.374696  -4.551418 -14.407102  5.1928292   0.348 0.775
 [8,]   0.8  7.708862  5.545448   2.163415  -8.824631  9.6404128   0.775 0.775
 [9,]   0.9  6.201298  2.407586   3.793712  -4.813915 10.4323243   0.362 0.775
 \end{verbatim}
\end{table}

\begin{figure*}[!ht]
\centering
\includegraphics[width=1\textwidth]{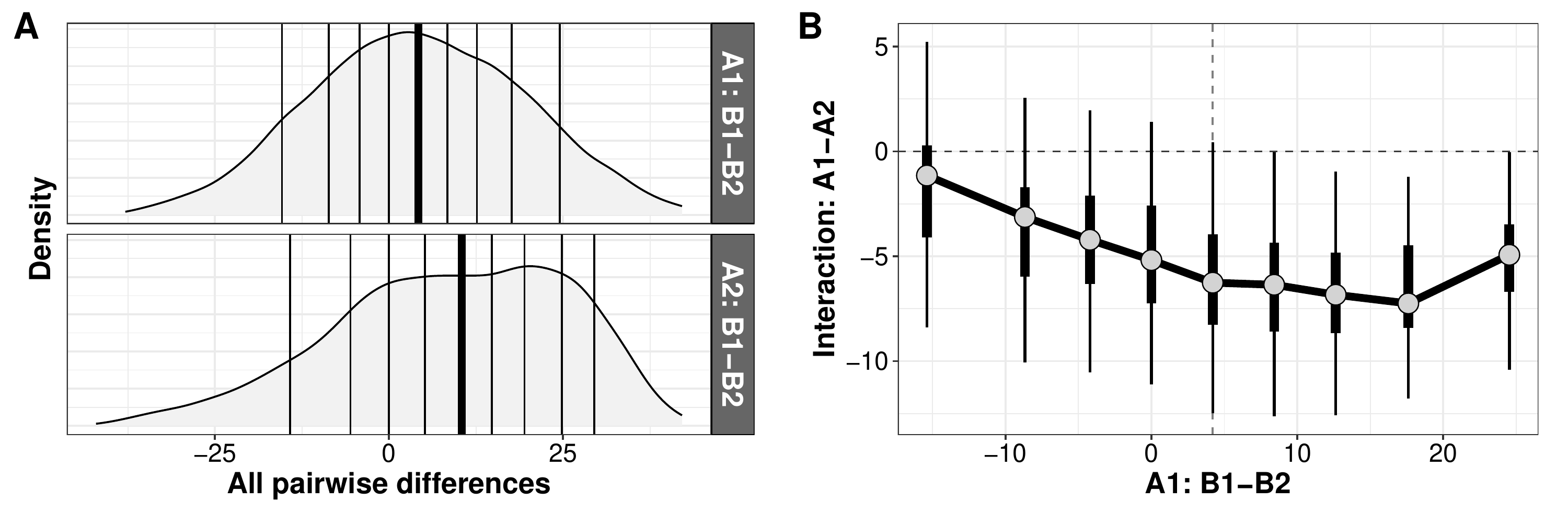} 
\caption{
\begin{small}
\textbf{Example: comparison of the deciles of all pairwise differences of perceived health.} (\textbf{A}) Distributions of all pairwise differences between participants with lower and higher CESD scores. Participants with lower levels of education are shown at the top, participants with higher levels at the bottom. The vertical lines mark the deciles, with a thicker line for the median. (\textbf{B}). Interaction plot; see details in Figure \ref{fig:ex_dec} caption.
\end{small}
}
\label{fig:ex_apd}
\end{figure*}

When testing (\ref{eq:apd}), plots of the difference scores help provide 
perspective. Figure \ref{fig:ex_apd}A illustrates this point. For each level of education (A1 = lower level; A2 = higher level), every participant with a lower CESD score was compared to every participant with a higher CESD score. As previously noted, when two distributions are identical,
the distribution of $D$ is symmetric about zero. Figure \ref{fig:ex_apd}A suggests 
that for both groups the distribution of the difference scores is shifted to the right, but with a stronger shift for group 2 (completed high school -- lower panel in Figure \ref{fig:ex_apd}A). A positive difference indicates higher perceived health in not depressed participants relative to depressed participants. For the second group, testing the hypothesis that the median of the difference scores is zero, the estimate is 10.5, with a [4.2, 14.7] 95\% confidence interval, $p=0$. In the first group the median is 4.21 [0, 8.4], $p=0.0765$.  

Table 2 summarizes the results when testing (\ref{eq:apd}).
Here, the .1, .25, .5, .75 and .9 quantiles are used, which is the default for the
R function \verb|iband| that was used. Now the unadjusted p-values indicate an
interaction in the upper tails of the two distributions. For example, the estimates of the 0.75 quantile indicate that for the first level of the first factor (did not complete high school), when comparing the not depressed group to the depressed group, there is 25\% chance of  getting a difference between perceived health values greater than 14.75, while for the second group there is  25\% chance of getting a difference greater than 22. But using the Hochberg adjustment shown here, no significant difference is found
at the .05 level and this remains the case when using the Benjamini--Hochberg method.

\begin{table}
\center
\caption{Result for Perceived health When Testing (\ref{eq:apd})}
\begin{verbatim}
     Quant  Est.Lev 1  Est.Lev 2       Dif    ci.low      ci.up p-value p.adj
[1,]  0.10 -15.380695 -14.229966 -1.150729  -8.58284  4.6709548   0.607 0.607
[2,]  0.25  -6.342360  -2.168466 -4.173895 -10.51543  2.0383410   0.165 0.330
[3,]  0.50   4.206226  10.466616 -6.260390 -12.44074  0.1001971   0.058 0.174
[4,]  0.75  14.746649  22.041140 -7.294491 -12.55662 -1.5587962   0.013 0.065
[5,]  0.90  24.529027  29.460025 -4.930998 -10.51350 -0.1045500   0.039 0.156
 \end{verbatim}
\end{table}

 
     






\section{Concluding Remarks}

The two new methods presented in this article help gain a deeper understanding of where and by how much groups differ in a 2x2 factorial design. With a sample size of at least 30 per group, all indications are that the methods perform well even when dealing with distributions that have a relatively high amount of skewness and kurtosis. These methods are implemented in the functions \verb|decinter| to test (\ref{eq:inter}) and \verb|iband| to test (\ref{eq:apd}). The R functions \verb|decinter| and \verb|iband| can be found in the file \verb|Rallfun-v41.txt|, downloadable from https://osf.io/nvd59/quickfiles. These functions, the R code for the simulation and the figures are available in the reproducibility package for the article (Wilcox \& Rousselet, 2023) and on GitHub (Rousselet, 2023). The reproducibility package also contains Rcpp code that is much faster to execute than the base R version (see notebook \verb|examples.Rmd|). By default, these functions use 2000 bootstrap samples and correct for multiple comparison using the FDR correction from Benjamini \& Hochberg (1995), which outperformed the FWER correction of Hochberg (1988) in our simulations. Even with the FDR correction, the two methods presented here remain conservative, so it would be worthwhile to explore other correction strategies. We have already considered several alternative methods, but they do not improve matters (Benjamini \& Yekutieli, 2001; Benjamini et al., 2006; Blanchard \& Roquain, 2008). Preliminary investigations of yet other approaches to control the FWER, such as using a maximum statistic distribution (Nichols \& Holmes, 2002), have not revealed any method that could significantly improve power in all situations, making recommendations difficult. While this issue requires further work, for the moment we recommend to use the FDR correction from Benjamini \& Hochberg (1995) by default.

More generally, the obvious concern with comparing multiple quantiles is that power might be negatively impacted due to controlling the FWER. But power might also be negatively impacted when focusing on a single quantile simply because other quantiles are being ignored. The example in the previous section demonstrated the risk of drawing conclusions from a single measure of central tendency or quantile. It is also worth keeping in mind that there is no free lunch in inference: methods that can reveal more complex patterns in the data, such as those proposed here, necessarily require larger sample sizes to reveal where and by how much distributions differ. For applications to the deciles considered here, given the increased uncertainty associated with the estimation of the 1st and 9th deciles, we recommend sample sizes of at least 30, echoing earlier recommendations (Wilcox et al., 2014).

Several strategies are worth considering to boost power, starting with testing more specific hypotheses involving only a subset of quantiles. For instance, one could imagine a cross-validation approach in which a large dataset is split between a discovery set and a testing set. Instead of testing each quantile individually, one could also look for specific patterns across quantiles, such as stochastic dominance, which is characterised by all quantile differences having the same sign, or differences in spread, which are characterised by monotonic trends across quantile differences (Rousselet et al. 2017). Other obvious strategies are to consider different quantile estimators and bootstrap methods. Preliminary investigations suggest that using the quantile(type=8) estimator recommended by Hyndman \& Fan (1996) improves type I error rates and power relative to the quantile(type=7) estimator for instance, but is still outperformed by the Harrell--Davis estimator in all situations. As for parametric methods, Goldman \& Kaplan (2018) have proposed a fast and powerful extension of the Kolmogorov--Smirnov test to compare all the quantiles of two independent groups (Kolmogorov, 1992; Stephens, 1992). However, their approach assumes no tied values, and it is unclear how it could be generalised to deal with interactions. 

A bootstrap approach provides enough flexibility to deal with a variety of experimental designs. As such, this work could be extended to mixed designs with within- and between-subject factors. A covariate could be handled using different strategies, including non-parametric methods with smoothers (Wilcox, 1997; Wilcox, 2021, chapter 12).

A referee inquired about how applied researchers might report  results based on the methods studied here. We suggest reporting
results as done in Tables 1 and 2.

\section{References}


Akinshin, A. (2022). Trimmed Harrell-Davis quantile estimator based
on the highest density interval of the given width.  {\em Communications in Statistics - Simulation and
Computation}. Online.  DOI: 10.1080/03610918.2022.2050396

Benjamini, Y. \& Hochberg, Y. (1995) Controlling the False Discovery Rate: A Practical and Powerful Approach to Multiple Testing. Journal of the Royal Statistical Society: Series B (Methodological), 57, 289–300.

Benjamini, Y., Krieger, A.M., \& Yekutieli, D. (2006) Adaptive Linear Step-up Procedures That Control the False Discovery Rate. Biometrika, 93, 491–507.

Benjamini, Y. \& Yekutieli, D. (2001) The control of the false discovery rate in multiple testing under dependency. The Annals of Statistics, 29, 1165–1188.

Blanchard, G. \& Roquain, E. (2008) Two simple sufficient conditions for FDR control. Electronic Journal of Statistics, 2, 963–992.

Bradley, J. V. (1978) Robustness? {\em British Journal of Mathematical and Statistical Psychology, 31},
 144--152. doi.org/10.1111/j.2044-8317.1978.tb00581.x

Cain, M., Zhang, Z. \& Yuan, K.-H. (2017). Univariate and multivariate skewness and kurtosis
for measuring nonnormality: Prevalence, influence
and estimation.     {\em Behavioral  Research, 49}, 1716--1735.   DOI 10.3758/s13428-016-0814-1.

Clark, F., Jackson, J., Carlson, M., Chou, C.-P.,  Cherry, B. J.,  Jordan-Marsh M.,  Knight, B. G.,  Mandel, D.,  Blanchard, J., Granger, D. A., Wilcox, R. R.,  Lai, M. Y.,  White, B.,  Hay, J.,  Lam, C., Marterella, A. \&  Azen, S. P. (2011). Effectiveness of a lifestyle intervention in promoting the well-being of independently living older people: results of the Well Elderly 2 Randomised Controlled Trial. {\em Journal of Epidemiology and Community Health, 66}, 782--790. doi:10.1136/jech.2009.099754.

De Neve, J. \& Thas, O. (2017).  A Mann--Whitney type effect measure of interaction for factorial designs.
{\em Communications in Statistics - Theory and Methods, 46}, 
  11243--11260.
   DOI: 10.1080/03610926.2016.1263739.

Doksum, K. A. \& Sievers, G. L. (1976). Plotting with confidence: graphical
 comparisons of two populations. {\em Biometrika, 63}, 421--434.
 doi.org/10.2307/2335720.

Eddelbuettel, D. \& Francois, R. (2011) Rcpp: Seamless R and C++ Integration. J. Stat. Soft., 40, 1–18.
 
Gao, X. \& Alvo, M. (2005). A nonparametric test for interaction in two-way layouts. {\em Canadian Journal of Statistics, 33}, 529--543.

Goldman, M. \& Kaplan, D. M. (2018). Comparing distributions by multiple testing across quantiles or CDF values. 
 {\em Journal of Econometrics, 206}, 143--166. 

Harrell, F. E. \& Davis, C. E. (1982). A new distribution-free quantile estimator. 
{\em Biometrika, 69}, 635--640. https://doi.org/10.1093/biomet/69.3.635.

Hoaglin, D.C. (1985) Summarizing Shape Numerically: The g-and-h Distributions. In Exploring Data Tables, Trends, and Shapes. John Wiley \& Sons, Ltd, pp. 461–513.

Hochberg, Y. (1988) A sharper Bonferroni procedure for multiple tests of significance. Biometrika, 75, 800–802.

Hommel, G. (1988). A stagewise rejective multiple test procedure based on a modified Bonferroni
test. {\em Biometrika,75} , 383--386.. doi.org/10.2307/2336190 

Hyndman, R. J. \& Fan, Y. (1996) Sample quantiles in statistical packages, American Statistician 50, 361–365. doi: 10.2307/2684934.

Koenker, R. \& Bassett, G. (1978). Regression quantiles. {\em Econometrika, 46},
 33--50. doi.org/10.2307/1913643 

Kolmogorov, A. (1992) On the Empirical Determination of a Distribution Function. In Kotz, S. \& Johnson, N.L. (eds), Breakthroughs in Statistics: Methodology and Distribution, Springer Series in Statistics. Springer, New York, NY, pp. 106–113.

Liu,  X., Song, Y. \& Zhang, K. (2022). An exact bootstrap-based
bandwidth selection rule for kernel quantile estimators.{\em  Communications in Statistics - Simulation
and Computation}, online.  DOI: 10.1080/03610918.2022.2110595

Lombard, F. (2005). Nonparametric confidence bands for a quantile comparison
 function. {\em Technometrics, 47}, 364--369.

Nichols, T.E. \& Holmes, A.P. (2002) Nonparametric permutation tests for functional neuroimaging: A primer with examples. Human Brain Mapping, 15, 1–25.
 
Navruz, G. \&  \"{O}zdemir, A. F. (2020). A new quantile estimator with weights based on a subsampling approach.
{\em British Journal of Mathematical and Statistical Psychology, 73}, 506--521.
https://doi.org/10.1111/bmsp.12198

Patel, K. M. \& Hoel, D. G. (1973). A nonparametric test for interaction in 
factorial experiments. {\em Journal of the American Statistical Association},
{\em 68}, 615--620. https://doi.org/10.2307/2284788

Rousselet, G. A. (2023) Code for article A Quantile Shift Approach To Main Effects And Interactions In A 2-By-2 Design. GitHub. https://github.com/GRousselet/decinter

Rousselet, G.A., Pernet, C.R., \& Wilcox, R.R. (2017) Beyond differences in means: robust graphical methods to compare two groups in neuroscience. European Journal of Neuroscience, 46, 1738–1748.

Stephens, M.A. (1992) Introduction to Kolmogorov (1933) On the Empirical Determination of a Distribution. In Kotz, S. \& Johnson, N.L. (eds), Breakthroughs in Statistics: Methodology and Distribution, Springer Series in Statistics. Springer, New York, NY, pp. 93–105.

Wickham, H. (2016) Ggplot2: Elegant Graphics for Data Analysis, 2nd edn, Use R! Springer International Publishing.

Wilcox, R.R. (1995) Comparing Two Independent Groups Via Multiple Quantiles. Journal of the Royal Statistical Society. Series D (The Statistician), 44, 91–99.

Wilcox, R.R. (1997) ANCOVA based on comparing a robust measure of location at empirically determined design points. British Journal of Mathematical and Statistical Psychology, 50, 93–103.

Wilcox, R. (2022).  {\em Introduction to Robust Estimation and Hypothesis Testing} 5th Edition. San Diego, CA: Academic Press

Wilcox, R. R.,  Erceg-Hurn, D., Clark, F. \& Carlson, M. (2014).  Comparing two  independent groups via the lower and upper quantiles. 
{\em  Journal of Statistical   Computation and Simulation, 84}, 1543--1551. 
 DOI: 10.1080/00949655.2012.754026
 
Wilcox, R. R. \& Rousselet, G. A. (2023). A quantile shift approach to main effects and interactions in a 2-by-2 design reproducibility package. figshare. Online resource.  https://doi.org/10.6084/m9.figshare.22927520.v1
 
Wilke, C.O. (2017) cowplot: Streamlined Plot Theme and Plot Annotations for “ggplot2.”

 \end{document}